\documentclass[submission,copyright,creativecommons]{eptcs}

\usepackage{newlfont}
\usepackage[below]{placeins}
\usepackage{flafter}
\usepackage{amsfonts}
\usepackage{amsmath}
\usepackage{amssymb}
\usepackage[american]{babel}
\usepackage{graphicx}
\usepackage{multirow}
\usepackage{url}
\usepackage[linesnumbered,ruled,vlined]{algorithm2e}
\usepackage{color}
\usepackage{mathtools}

\providecommand{\event}{QSQW 2020} 
\usepackage{underscore}           

\allowdisplaybreaks

\newcommand{\ket}[1]{\big| #1 \big\rangle}
\newcommand{\bra}[1]{\big\langle #1 \big|}
\newcommand{\braket}[2]{\big\langle #1 \big| #2 \big\rangle}                 
\newcommand{\bracket}[3]{\big\langle #1 \big| #2 \big| #3 \big\rangle}       

\newcommand{\e}{\textrm{e}}

\newcommand{\qed}{\nobreak \ifvmode \relax \else
      \ifdim\lastskip<1.5em \hskip-\lastskip
      \hskip1.5em plus0em minus0.5em \fi \nobreak
      \vrule height0.75em width0.5em depth0.25em\fi}

\title{Discrete-Time Quantum Walks on Oriented Graphs}
\author{Bruno Chagas
\institute{Federal University of Minas Gerais -- UFMG, \\Belo Horizonte, MG, Brazil}
\email{brunomcct@gmail.com}
\and
Renato Portugal
\institute{National Laboratory of Scientific Computing -- LNCC, \\Petr\'{o}polis, RJ,  25651-075, Brazil}
\email{portugal@lncc.br}
}
\def\titlerunning{DTQW on Oriented Graphs}
\def\authorrunning{B. Chagas and R. Portugal}

\begin{document}
\maketitle

\begin{abstract}
The interest in quantum walks has been steadily increasing during the last two decades. It is still worth to present new forms of quantum walks that might find practical applications and new physical behaviors. In this work, we define discrete-time quantum walks on arbitrary oriented graphs by partitioning a graph into tessellations, which is a collection of disjoint cliques that cover the vertex set. By using the adjacency matrices associated with the tessellations, we define local unitary operators, whose product is the evolution operator of our quantum walk model. We introduce a parameter, called $\alpha$, that quantifies the amount of orientation. We show that the parameter $\alpha$ can be tuned in order to increase the amount of quantum walk-based transport on oriented graphs.

\

\end{abstract}

\section{Introduction}

Quantum walk is an active research area due to its ability to simulate complex quantum systems~\cite{BN16,AMAMPPDAD19} and its usefulness to build new quantum algorithms~\cite{Por18book}. There are many versions of quantum walks, which must obey some basic rules, such as, the position of the walker must be modeled by a graph, or some discrete spatial structure, and the evolution operator itself must be local or is the product of local operators, that is, the dynamics must obey a property that some authors call \textit{graph locality}~\cite{TLSPBK16,AST18}. The definition of local operators (in the context of quantum walks) relies on the graph structure and can be colloquially stated in the following way: An operator is local if its action on a walker that is on vertex $v$ shifts the walker to the neighborhood of $v$  before evolving the walker's state to vertices beyond the neighborhood of $v$. In the discrete-time case, the walker moves to the neighborhood of the neighborhood of $v$ only after the action of at least two local operators, which are different in general, because the walker stays trapped in a subgraph if we apply the same local operator repeatedly. 
In any case, the dynamics of the quantum walk is a subcase of quantum dynamics, which is not constrained by graph structures.

One of the earliest attempts to define quantum walks on \textit{directed graphs} was presented by Severini~\cite{Sev03,Sev02}, who defined the concept of a digraph of a unitary matrix and established the conditions that the digraph must fulfill to be a digraph of a unitary matrix. He has also pointed out that the walker of the coined model on a simple graph $\Gamma$ proposed in~\cite{AAKV01} steps on the directed arcs of the complete digraph, whose underlying graph is $\Gamma$. Ref.~\cite{Mon07} has further explored the definition of quantum walks on directed graphs by characterizing the notion of graph reversibility. Ref.~\cite{HM09} has used the concept of quantum walk on directed graphs to enhance quantum transport on the line.  Ref.~\cite{TLSPBK16} has analyzed continuous-time quantum walk on directed bipartite graphs with the goal of suppressing quantum transport.

In this work, we propose a definition of a discrete-time quantum walk on oriented graphs (a directed graph with no bidirected edges). In the continuous-time model, the most natural route to define quantum walks on oriented graphs is to use the generalized adjacency matrix $(zA-z^*A^T)$, where $A$ is the original adjacency matrix and $z$ is a complex number~\cite{GM17}, as the model's Hamiltonian $H$ and evolution operator $U=\exp(itH)$. The problem in discrete-time models is that $U$ is nonlocal in general. We describe an escape route by partitioning the vertex set into \textit{cliques}\footnote{A clique is a subset of vertices that induces a complete subgraph.} and by using a sequence of adjacency matrices $A_1$, $A_2$, etc.
The evolution operator in this case is the product of local unitary operators $U_1$, $U_2$, etc., as is unavoidable in the discrete-time case, where each local unitary $U_k$ is $\exp(i\theta H_k)$, where $H_k=\exp(i \alpha)A_k-\exp(-i \alpha)A_k^T$, where $\alpha$ and $\theta$ are real parameters. The fact that each element of the graph partitioning is a clique guarantees that $U_k$ is local. The technique of partitioning the vertex set into cliques used in this context is the same technique used to define staggered quantum walks~\cite{PSFG16,POM17}.

After defining the evolution operator of discrete-time quantum walks on oriented graphs, we analyze quantum walks on oriented lines and oriented two-dimensional lattices with the focus on quantum transport. Since parameter $\theta$ plays the same role of the namesake parameter in the staggered model, we focus on parameter $\alpha$, which tunes the amount of graph orientability: If $\alpha=0$, no orientation is taken and if $\alpha=\pi$, the orientation is maximal. We show that quantum transport can be enhanced or decreased by tuning parameter $\alpha$. 

The structure of this paper is as follows. Sec.~\ref{sec:evol} describes how to obtain a local operator based on partitioning the vertex set into clique and shows why it is necessary at least two partitions. Sec.~\ref{sec:FB} describes the evolution operator of quantum walks on oriented graphs. Sec.~\ref{sec:SD} analyzes quantum walks on the oriented line and presents analytical results for the quantum transport.  Sec.~\ref{sec:grid2d} analyzes quantum walks on the oriented two-dimensional lattice. Sec.~\ref{sec:conc} presents our conclusions.


\section{A recipe to obtain the evolution operator}\label{sec:evol}

One of the main goals of studying quantum walks on a graph $\Gamma$ is to provide tools for building Hamiltonians that describe the time evolution of quantum systems, whose structure is based on $\Gamma$. We are interested in the class of quantum walks whose Hilbert space is spanned by the graph vertices, sometimes called coinless quantum walks. Since locality is the key issue here, it is important to define what is a local Hamiltonian in the quantum-walk context in order to distinguish from the term ``local'' used in quantum computing.~\footnote{If one implements the evolution operator of a quantum walk on a qubit-based quantum computer, the notion of locality refers not to the qubit connectivity graph of the quantum computer, but to the graph that describes the walker's position.} If the walker is on a vertex $v$, a local Hamiltonian is one that drives the walker to a vertex in the neighborhood of $v$. If $w$ is a vertex that does not belong to the neighborhood of $v$ or, let us say, $w$ is far from $v$, then the walker requires many steps (in the discrete-time case) or a long time (in the continuous-time case) to move from $v$ to $w$. If the time evolution is driven by only one local Hamiltonian $H$ (for instance, the adjacency matrix) and the unitary evolution operator is $U=\exp(i H t)$, then $t$ must be continuous because we have to use an infinitesimal time to counteract terms of order two or larger in the Taylor expansion of the exponential function, which are nonlocal terms~\cite{Por18book}. In the discrete-time case, on the other hand, we have to employ at least two local Hamiltonians alternately, that is, the state of the system at time $t$ is $\ket{\psi(t)}=U^t\ket{\psi(0)}$, where $\ket{\psi(0)}$ is the initial state, $U$ is the evolution operator that must be the product of local unitary operators $U=U_2U_1$, where $U_1=\exp(i\theta H_1)$, $U_2=\exp(i\theta H_2)$, and $H_1$, $H_2$ are local Hamiltonians, $\theta$ is an angle. The area of quantum walks must provide recipes that describe ways of obtaining $H_1$ and $H_2$. Let us review one of these recipes.

\begin{figure}[h!] 
\centering\vspace{0.0cm}(a)
\includegraphics[scale=1.2]{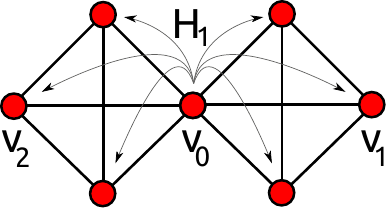}\hspace{1.5cm}(b)
\includegraphics[scale=1.2]{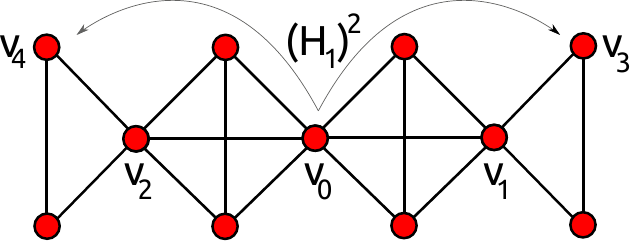}
\caption{Panel (a): Suppose that the walker is initially on vertex $v_0$, and there are two cliques incident to $v_0$ (the remaining of the graph is not shown in this panel). Usually, the action of Hamiltonian $H_1$ shifts the walker to both cliques, as shown by the arrows. Panel (b): If we repeat the action of $H_1$ again, the position of the walker spreads even further, for instance, to vertices $v_3$ and $v_4$ as shown. The distance between $v_3$ and $v_4$ is 4. This mean that $(H_1)^2$ is nonlocal because its action shifts the walker from $v_0$ to $v_3$. Vertices $v_0$ and $v_3$ are not neighbors; the distance is 2. Since $(H_1)^2$ is nonlocal in this case, $U_1=\exp(i H_1)$ is also nonlocal and cannot be used in the definition of the evolution operator of a discrete-time quantum walk. }
\label{fig:cliques}
\end{figure}

Let us focus our attention on how to obtain $H_1$ in the discrete-time case. The first problem we face is that $(H_1)^2$ is a term in the Taylor expansion of $\exp(i\theta H_1)$. If $(H_1)^2$ or any other higher order term is nonlocal then $U_1$ is nonlocal. Can  $(H_1)^2$ be local? To answer this question we need to understand what produces the nonlocality of $(H_1)^2$. The root of the problem lies in the fact that the action of $H_1$ on a vertex $v_0$ in general spreads the amplitudes of the wave function over vertices whose distance is 2 because it is possible to have vertices $v_1$ and $v_2$ in the neighborhood of $v_0$ so that the distance between $v_1$ and $v_2$ is 2 (see Fig.~\ref{fig:cliques}a). If $H_1$ spreads the wave function over vertices that have distance 2, then $(H_1)^2$ spreads the wave function over vertices whose distance is 4. Then, $(H_1)^2$ is nonlocal because there is a vertex $v_3$ in the neighborhood of $v_1$ or $v_2$ in the expression of $(H_1)^2\ket{v_0}$ so that the distance between $v_0$ and $v_3$ is 2 (see Fig.~\ref{fig:cliques}b). The same argument shows that higher order powers of $H_1$ are also nonlocal in general.

\begin{figure}[h!] 
\centering\vspace{0.0cm}
(a)\hspace{0.3cm}\includegraphics[scale=0.6]{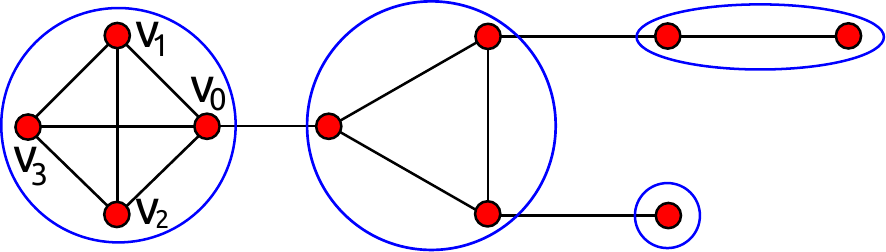}$\Gamma$\hspace{2cm}(b)
\includegraphics[scale=0.6]{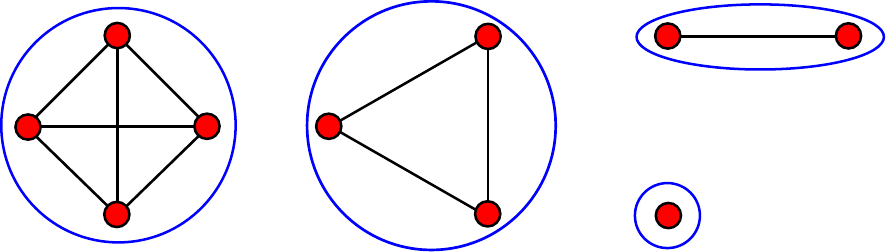}$\mathcal{T}_1$
\caption{(a) Example of a partition of graph $\Gamma$ into cliques. (b) Subgraph $\mathcal{T}_1$ obtained by deleting the edges of $\Gamma$ that do not belong the tiles of the partition. }
\label{fig:graph1}
\end{figure}

To escape this bad fate (nonlocal $U_1$), the action of $H_1$ must spread the wave function over vertices whose distance is 1, that is, if
\[H_1\ket{v_0}=\alpha\ket{v_0}+\beta\ket{v_1}+\gamma\ket{v_2}+\cdots\]
then the pairwise distance of vertices $v_0$, $v_1$, $v_2$, etc.~must be 1. Those vertices must be a \textit{clique}, that is $v_0$, $v_1$, $v_2$, etc.~must induce a complete subgraph of $\Gamma$. For instance, the set of vertices $\{v_0,v_1,v_2,v_3\}$ in Fig.~\ref{fig:graph1}a is a clique of size 4, and each encircled subsets of vertices are examples of cliques of sizes 1, 2, 3, and 4. Besides, for the sake of homogeneity,  we have to demand that the action of $H_1$ on a walker on \textit{any} vertex of the clique $\{v_0,v_1, v_2, \text{etc.}\}$ spread the position of the walker inside this clique. After these demands, the problem shown in Fig.~\ref{fig:cliques} does not arise, that is, $(H_1)^2$ is local.

To define $H_1$, we partition the set of vertices into cliques (called tiles), which defines a subgraph $\mathcal{T}_1$ after erasing the edges of $\Gamma$ that do not belong to the tiles (see Fig.~\ref{fig:graph1}b). $H_1$ is the adjacency matrix of $\mathcal{T}_1$, which is also denoted by $A(\mathcal{T}_1)$. Note that $(H_1)^k$ is local for any integer $k$ because the walker never leaves the neighborhood (there are no edges linking two different tiles). Then, $U_1$ is local.

\begin{figure}[h!] 
\centering\vspace{0.0cm}
(a)\hspace{0.3cm}\includegraphics[scale=0.6]{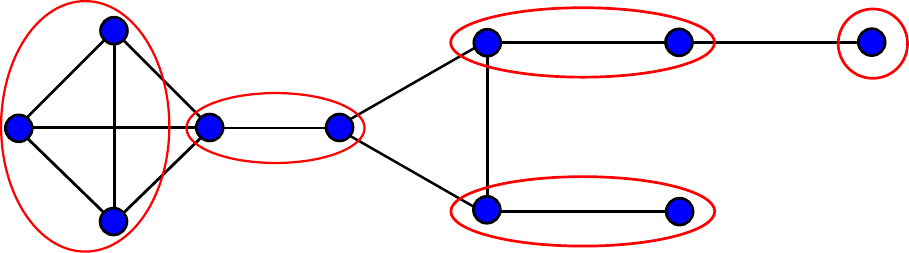}$\Gamma$\hspace{2cm}(b)
\includegraphics[scale=0.6]{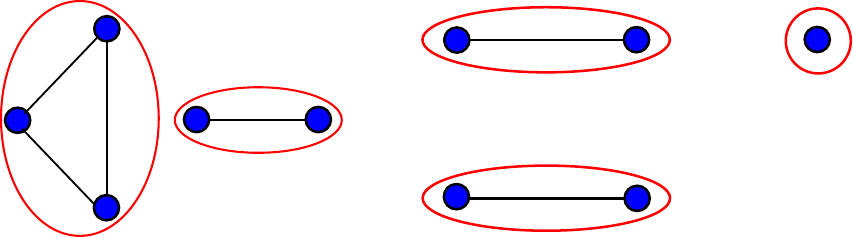}$\mathcal{T}_2$
\caption{Example of a second partition of the graph $\Gamma$ of Fig.~\ref{fig:graph1} into cliques. (b) Subgraph $\mathcal{T}_2$ obtained by deleting the edges of $\Gamma$ that do not belong the tiles of the second partition.}
\label{fig:graph2}
\end{figure}

Now we understand why in discrete-time models we need more than one local unitary operator. The recipe above generates a local unitary operator that traps the walker in tiles. The edges that do not belong to the tiles play no role in the dynamic driven by $U_1$. The way out is to obtain a second partition, as the one described in Fig.~\ref{fig:graph2}a, which defines a subgraph $\mathcal{T}_2$ induced by the new tiles (see Fig.~\ref{fig:graph2}b), and $H_2$ is the adjacency matrix of $\mathcal{T}_2$. The recipe for this example is complete and the evolution operator of the quantum walk is $U=U_2U_1$, where 
$H_1=A(\mathcal{T}_1)$, $H_2=A(\mathcal{T}_2)$, and
\begin{eqnarray*}
U_1=\e^{i\theta A(\mathcal{T}_1)},\nonumber\\
U_2=\e^{i\theta A(\mathcal{T}_2)}.
\end{eqnarray*}
In many cases, it is desirable to use Hamiltonians whose square is the identity operator. This can be accomplished by finding real numbers $a,b$ such that $H_1=aI+bA(\mathcal{T}_1)$ and $(H_1)^2=I$. This way of defining a discrete-time quantum walk is as close as possible to the continuous-time model.

The partition of a graph into cliques was studied in graph theory under the name of \textit{equivalence graph} and the union of the partitions under the name \textit{equivalence covering}~\cite{Alo86}. In the notation of the staggered quantum walk model, a partition is called by \textit{tessellation} and the union of the partitions by \textit{tessellation cover}~\cite{PSFG16,ACFKMPP20}. Note that there are graphs called $k$-tessellable for which it is necessary $k>2$ partitions to cover the graph edges, for instance, odd cycles. In the staggered model, there is an extra demand: $H_1$ and $H_2$ must be involutory, that is, $(H_1)^2=(H_2)^2=I$. In this case, $H_1$ and $H_2$ are Hermitian and unitary. The partitioning of graph into cliques generalizes in some sense the discussion presented by Meyer~\cite{Mey96} for the one-dimensional lattice.

\section{Oriented graphs}\label{sec:FB}

An oriented graph is a directed graph with no bidirected edges.
The adjacency matrix of an oriented graph is non-Hermitian~\cite{Die12}. This means that the recipe outlined in the previous section does not work because the exponentiation of a non-Hermitian operator generates a non-unitary operator. The way out is to consider both the oriented graph and its transpose (all arcs are reversed), also called converse. If $A$ is the adjacency matrix of an oriented graph $\vec \Gamma$ then $A^T$  is the adjacency matrix of the transpose of $\vec \Gamma$. Note that $A+A^T$ is the adjacency matrix of the symmetric directed graph $\Gamma$ obtained from $\vec \Gamma$ by converting all arcs into bidirected edges. An interesting Hamiltonian associated with both $\vec \Gamma$ and its transpose is
\[H=\e^{i\alpha}A+\e^{-i\alpha}A^T,\]
where $\alpha$ is an angle. In this case, the operator $U=\exp(iH)$ is unitary but unfortunately nonlocal in the general case. It is acceptable to use $\exp(iHt)$ in the context of continuous-time quantum walks~\cite{Luetal16,GL20}. 
In the discrete-time case, we have to obtain a tessellation cover of the symmetric directed graph $\Gamma$, which induces partitions of both $\vec \Gamma$ and its transpose. Suppose that $\Gamma$ is 2-tessellable and let $A_1$, $A_2$ be the adjacency matrices of the oriented subgraphs of $\vec \Gamma$ induced by the tessellations. We define
\begin{eqnarray*}
H_1&=&{\e^{i\alpha}\,A_1\,+\,\e^{-i\alpha}A_1^T},\nonumber\\
H_2&=&{\e^{i\alpha}\,A_2\,+\,\e^{-i\alpha}A_2^T}.
\end{eqnarray*}
The corresponding local unitary operators are
\begin{eqnarray*}
U_1&=&\e^{i\theta H_1},\nonumber\\
U_2&=&\e^{i\theta H_2},
\end{eqnarray*}
and the resulting evolution operator is $U=U_2U_1$. 

From the viewpoint of graph theory, matrix $(A-A^T)$ is called skew symmetric adjacency matrix of an oriented graph~\cite{CCFGHKMT12} and matrix $A'$, where $A'_{k\ell}=-A'_{\ell k}=i$ if there is an oriented arc from $v_k$ to $v_\ell$ and $A'_{k\ell}=A'_{\ell k}=1$ if there is an edge linking $v_k$ and $v_\ell$, is called the Hermitian-adjacency matrix of mixed graphs (directed graphs with arcs and edges)~\cite{LL15,GM17}. It natural to consider the generalized adjacency matrix $(zA-z^*A^T)$ with $z=\theta\e^{i\alpha}\in \mathbb{C}$ for oriented graphs, which is also Hermitian. In the context of this interpretation, we say that we have defined a discrete-time quantum walk on an oriented graph.

\section{Oriented line}\label{sec:SD}

\begin{figure}[!h] 
\centering
\vspace{0.2cm}
\includegraphics[scale=0.6]{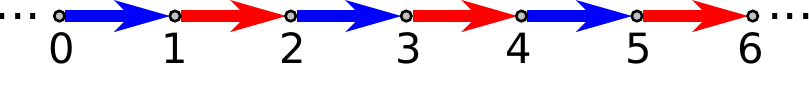}\hspace{0.2cm}$\e^{+i\alpha}$\vspace{0.2cm} \\
\includegraphics[scale=0.6]{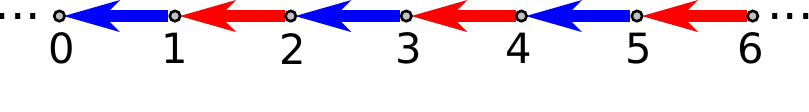}\hspace{0.2cm} $\e^{-i\alpha}$
\caption{An oriented line with weight $\e^{i\alpha}$ and its transpose with weight $\e^{-i\alpha}$. The colors specify to which tessellation the arcs belong. }
\label{fig:graph2_line1}
\end{figure}

\begin{figure}[!h] 
\centering\vspace{0.0cm}\vspace{0.2cm}
\includegraphics[scale=0.6]{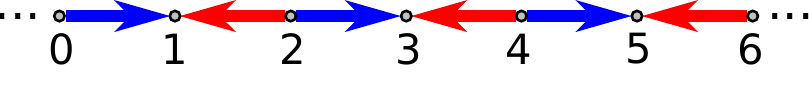}\hspace{0.2cm} $\e^{+i\alpha}$\vspace{0.2cm} \\
\includegraphics[scale=0.6]{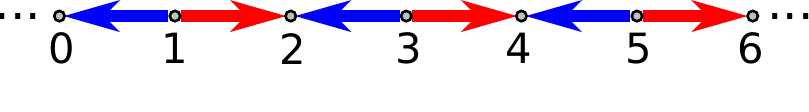}\hspace{0.2cm} $\e^{-i\alpha}$
\caption{A second form of orientation of the line. }
\label{fig:graph2_line2}
\end{figure}

As first examples, let us consider two cases of oriented line, as shown in Fig.~\ref{fig:graph2_line1} and~\ref{fig:graph2_line2}. The adjacency matrices of the tessellations of the right-hand oriented line are
\begin{eqnarray}
A^+=\sum_{x}\ket{2x}\bra{2x+1},
\end{eqnarray}
and 
\begin{eqnarray}
A^-=\sum_{x}\ket{2x-1}\bra{2x},
\end{eqnarray}
where $A^+$ refers to the blue tessellation and $A^-$ to the red tessellation.

The calculations are simpler if we use the Fourier basis, which is given by~\cite{PBF15}
\begin{eqnarray}
   \ket{\psi_{k}^0} &=& \sum_{x=-\infty}^\infty \textrm{e}^{-2xki}\ket{2x},\\
   \ket{\psi_{k}^1} &=& \sum_{x=-\infty}^\infty \textrm{e}^{-(2x+1)ki}\ket{2x+1},
\end{eqnarray}
where $k\in [-\pi,\pi]$. For a fixed value of $k$, those vectors define a subspace of the Hilbert space that is invariant under the action of the adjacency matrices, that is
\begin{equation}
A^\pm\ket{\psi_{k}^j}=\sum_{j'=0}^1\bracket{j'}{a^\pm}{j}\ket{\psi_{k}^{j'}},
\end{equation}
where $j\in\{0,1\}$ and $a^\pm$ are 2-by-2 matrices associated with $A^\pm$, given by
\begin{eqnarray}
a^+=
\left[\begin{array}{cc}
 0 & \e^{- ik} \\
  0 &  0
\end{array}\right],
\end{eqnarray}
and
\begin{eqnarray}
a^-=
\left[\begin{array}{cc}
 0 &  0 \\
  \e^{- ik} &  0
\end{array}\right].
\end{eqnarray}
The reduced versions of the local operators associated with the tessellations are
\begin{eqnarray}
u^\pm &=&\e^{i\theta\left(\e^{i\alpha}a^\pm+\e^{-i\alpha}(a^\pm)^\dagger\right)},
\end{eqnarray}
and the reduced version of the evolution operator is $$u=u^-u^+,$$  
which is given by
\begin{equation}
u=\left[\begin{array}{cc}
 \cos^2\theta -\sin^2\theta\,\,\textrm{e}^{2 i (k-\alpha)} & 
  i \sin 2\theta\,\cos (k-\alpha) \\
  i \sin 2\theta\,\cos (k-\alpha) &  
 \cos^2\theta -\sin^2\theta\,\,\textrm{e}^{2 i (\alpha-k)}
\end{array}\right].
\end{equation}
The eigenvalues of $u$ are $\textrm{e}^{\pm  i  \lambda}$, where
\begin{equation}\label{eq:costheta}
\cos\lambda=\cos^2\theta - \sin^2\theta\,\cos 2(k-\alpha).
\end{equation}
The non-trivial normalized eigenvectors of $u$ are 
\begin{equation}
\ket{v_{k}^{\pm}}=\frac{1}{\sqrt{C^\pm}}\left(\begin{array}{c}
\sin 2\theta \,\cos (k-\alpha)\\
\sin^2\theta\, \sin 2(k-\alpha) \pm \sin\lambda
\end{array}\right),\label{eq:eigenvec}
\end{equation}
where
\begin{equation}
C^\pm=2\,\sin\lambda\,(\sin\lambda \pm \sin^2\theta\,\sin 2(k-\alpha)).
\end{equation}

The evolution operator acting on the Hilbert space spanned by the vertices is
\begin{equation}
U=\sum_{k}\sum_{j,j'=0}^1\bracket{j'}{u}{j}\,\ket{\psi_{k}^{j'}}\bra{\psi_{k}^j}.
\end{equation}
The normalized eigenvectors of $U$ associated with eigenvalues $\textrm{e}^{\pm  i  \lambda}$ are 
\begin{eqnarray}
\ket{V_{k}^{\pm}}&=&\frac{1}{\sqrt{C^\pm}}\left(\sin 2\theta \cos (k-\alpha)\,\ket{\psi_{k}^0}+(\sin^2\theta \sin 2(k-\alpha) \pm \sin\lambda)\ket{\psi_{k}^1}\right).
\label{eq:Uev}
\end{eqnarray}
The evolution operator analyzed in Ref.~\cite{PBF15} is obtained if we take $\alpha=0$ and $\theta=\pi/2$ and the one analyzed in Ref.~\cite{POM17} is obtained if we take $\alpha=0$.

Let us focus on the case described in Fig.~\ref{fig:graph2_line2}. The adjacency matrices are
\begin{eqnarray}
A^\pm =\sum_{x}\ket{2x}\bra{2x\pm 1}.
\end{eqnarray}
The reduced adjacency matrices are
\begin{eqnarray}
a^\pm=
\left[\begin{array}{cc}
 0 & \e^{\mp ik} \\
  0 &  0
\end{array}\right],
\end{eqnarray}
and the reduced version of the evolution operator is 
\begin{equation}
u=\left[\begin{array}{cc}
 \cos^2\theta -\sin^2\theta\,\,\textrm{e}^{2 i k} & 
  i \sin 2\theta\,\cos k\,\,\textrm{e}^{i\alpha} \\
  i \sin 2\theta\,\cos k\,\,\textrm{e}^{-i\alpha} &  
 \cos^2\theta -\sin^2\theta\,\,\textrm{e}^{-2 i k}
\end{array}\right].
\end{equation}
The eigenvalues of $u$ are $\textrm{e}^{\pm  i  \lambda}$, where
\begin{equation}\label{eq:costheta2}
\cos\lambda=\cos^2\theta - \sin^2\theta\,\cos 2 k .
\end{equation}
The non-trivial normalized eigenvectors of $u$ are 
\begin{equation}
\ket{v_{k}^{\pm}}=\frac{1}{\sqrt{C^\pm}}\left(\begin{array}{c}
\sin 2\theta \,\cos  k\,\,\textrm{e}^{i\alpha} \\
\sin^2\theta\, \sin 2 k  \pm \sin\lambda
\end{array}\right),\label{eq:eigenvec2}
\end{equation}
where
\begin{equation}
C^\pm=2\,\sin\lambda\,(\sin\lambda \pm \sin^2\theta\,\sin 2 k ).
\end{equation}

\subsection{Transport}

One of the main transport measures is the first moment of the quantum walk, which specifies the motion of the wave function. It is also important to have a standard deviation as small as possible. In this section, we find what are the optimal values of $\alpha$ that maximizes the first moment and minimizes the standard deviation. The optimal value of $\alpha$ usually depends on the initial condition, which we take localized at most in two sites. 

The first moment is defined as
\begin{equation}
 \langle x \rangle = \sum_{x=-\infty}^\infty x \,\left|\braket{x}{\psi(t)}\right|^2,
\end{equation}
where $\ket{\psi(t)}$ is the state of the quantum walk after $t$ steps. To calculate  $\langle x \rangle$, we need to find the first derivative at $k=0$ of the characteristic function $\varphi_X(k)$, which is the expected value of $\textrm{e}^{\textrm{i}k X}$, where $X$ is the position operator, that is, $\varphi_X(k)=\bracket{\psi(t)}{\textrm{e}^{{i}k X}}{\psi(t)}$. 
Using the methods outlined in~\cite{Kon02} or ~\cite{Por18book,SPB15}, we obtain
\begin{equation}
\frac{\langle x \rangle}{t} = 
 2 \,\big(  \left| a \right|^2 - \left| b \right|^2  \big)  \left( 1-\cos 
\theta \right) + 
{\frac {i\sin \left( 2\,\theta \right) \left( \bar{a}\,b\,{{e}^{i\alpha}}-a\,\bar{b}\,{{e}^{-i\alpha}} \right) }{1+|\cos \theta\, | }}+O\left(\frac{1}{t}\right)
\end{equation}
for the initial condition is $\ket{\psi(0)}=a\ket{0}+b\ket{1}$. Similar calculations show that 
\begin{equation}
\frac{\langle x^2 \rangle}{t^2} = 
 4 \,  \left( 1-|\cos\theta| \right) 
+O\left(\frac{1}{t}\right)
\end{equation}
the even moments do not depend on $\alpha$ (they do depend on $\theta$). For $a=b=1/\sqrt{2}$, the maximum transport is obtained when $\alpha=\pi/2$ and $\cos \theta=(\sqrt{5}-1)/2$. In this case $\langle x \rangle\approx 0.60\,t$ asymptotically, which is the maximum possible value for this initial condition and the standard deviation $\sigma=\sqrt{\langle x^2 \rangle-\langle x \rangle^2}$ is minimum.

\section{Oriented two-dimensional lattice}\label{sec:grid2d}

\begin{figure}[!h] 
\centering\vspace{0.0cm}\vspace{0.2cm}
\includegraphics[scale=0.7]{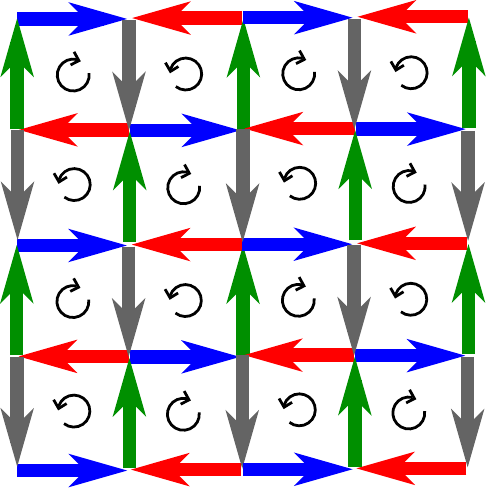}\hspace{0.7cm}\includegraphics[scale=0.7]{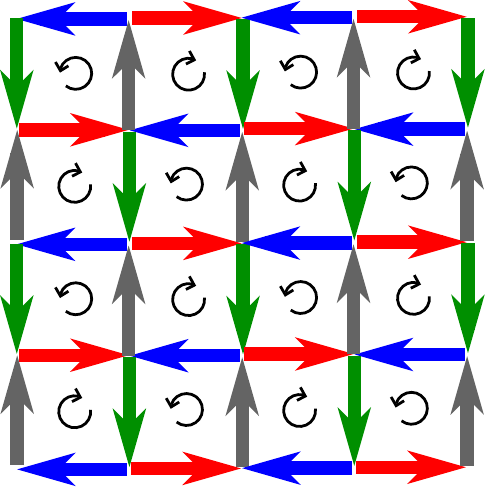}\\ {\centering $\e^{i\alpha}$ \hspace{3.0cm} $\e^{-i\alpha}$}
\caption{An oriented two-dimensional lattice with weight $\e^{i\alpha}$ and its transpose with weight $\e^{-i\alpha}$. The colors specify to which tessellation the arcs belong. By identifying the vectors with the same color at the boundaries, we obtain a finite lattice with $N=16$ and cyclic boundary conditions.}
\label{fig:graph3_grid2D}
\end{figure}

Let us consider an oriented two-dimensional lattice as described by the left-hand graph of Fig.~\ref{fig:graph3_grid2D}. We consider two kinds: infinite lattices and finite lattices with cyclic boundary conditions, where $N=(2n)^2$ is the number of vertices, assuming that $n>1$. The Hilbert space is spanned by $\{\ket{x,y}:0\le x,y<2n\}$ in the finite case.

The adjacency matrices of the tessellations of the left-hand graph of Fig.~\ref{fig:graph3_grid2D} are
\begin{eqnarray}
A_x^\pm=\sum_{x+y=\text{even}}\ket{x,y}\bra{x\pm 1,y},\\
A_y^\pm=\sum_{x+y=\text{odd}}\ket{x,y}\bra{x,y\pm 1},
\end{eqnarray}
where $A_x^\pm$ refer to tessellations blue and red, respectively, and $A_y^\pm$ refer to tessellations green and gray, respectively. The adjacency matrices of the tessellations of the right-hand graph of Fig.~\ref{fig:graph3_grid2D} are the transpose of $A_x^\pm$ and $A_y^\pm$.

The calculations are simpler if we use the Fourier basis, which is given by~\cite{PF17}
\begin{eqnarray}
  \ket{\psi^{0}_{k\ell}}&=&\frac{1}{\sqrt{2}\,n}\sum_{x,y=0}^{n-1}\left( {\e}^{2x\tilde{k}+2y\tilde{\ell}}\ket{2x,2y} +{\e}^{(2x+1)\tilde{k}+(2y+1)\tilde{\ell}}\ket{2x+1,2y+1}\right),   \label{psi_0} \\
  \ket{\psi^{1}_{k\ell}}&=&\frac{1}{\sqrt{2}\,n}\sum_{x,y=0}^{n-1}\left({\e}^{2x\tilde{k}+(2y+1)\tilde{\ell}}\ket{2x,2y+1}  +{\e}^{(2x+1)\tilde{k}+2y\tilde{\ell}}\ket{2x+1,2y} \right), \label{psi_1}
\end{eqnarray}
where $\tilde{k}={\pi k}/{n}$, $\tilde{\ell}={\pi \ell}/{n}$, and variables $k,\ell$ run from 0 to $2n-1$. For fixed values of $k$ and $\ell$, those vectors define a plane that is invariant under the action of the adjacency matrices, that is
\begin{equation}
A_{x,y}^\pm\ket{\psi_{k\ell}^j}=\sum_{j'=0}^1\bracket{j'}{a_{x,y}^\pm}{j}\ket{\psi_{k\ell}^{j'}},
\end{equation}
where $j\in\{0,1\}$ and $a_{x,y}^\pm$ are 2-by-2 matrices associated with $A_{x,y}^\pm$, given by
\begin{eqnarray}
a_x^\pm=
\left[\begin{array}{cc}
 0 & 0 \\
  \e^{\mp i\tilde{k}} &  0
\end{array}\right]
\end{eqnarray}
and
\begin{eqnarray}
a_y^\pm=
\left[\begin{array}{cc}
 0 & \e^{\mp i\tilde{\ell}}\\
 0   &  0
\end{array}\right].
\end{eqnarray}
The reduced versions of the local operators associated with the tessellations are
\begin{eqnarray}
u_{x,y}^\pm &=&\e^{i\theta\left(\e^{i\alpha}a_{x,y}^\pm+\e^{-i\alpha}(a_{x,y}^\pm)^T\right)}
\end{eqnarray}
and the reduced version of the evolution operator is $$u=-u_y^-u_x^-u_y^+u_x^+.$$

\begin{figure}[ht!] 
\centering
\includegraphics[scale=0.085]{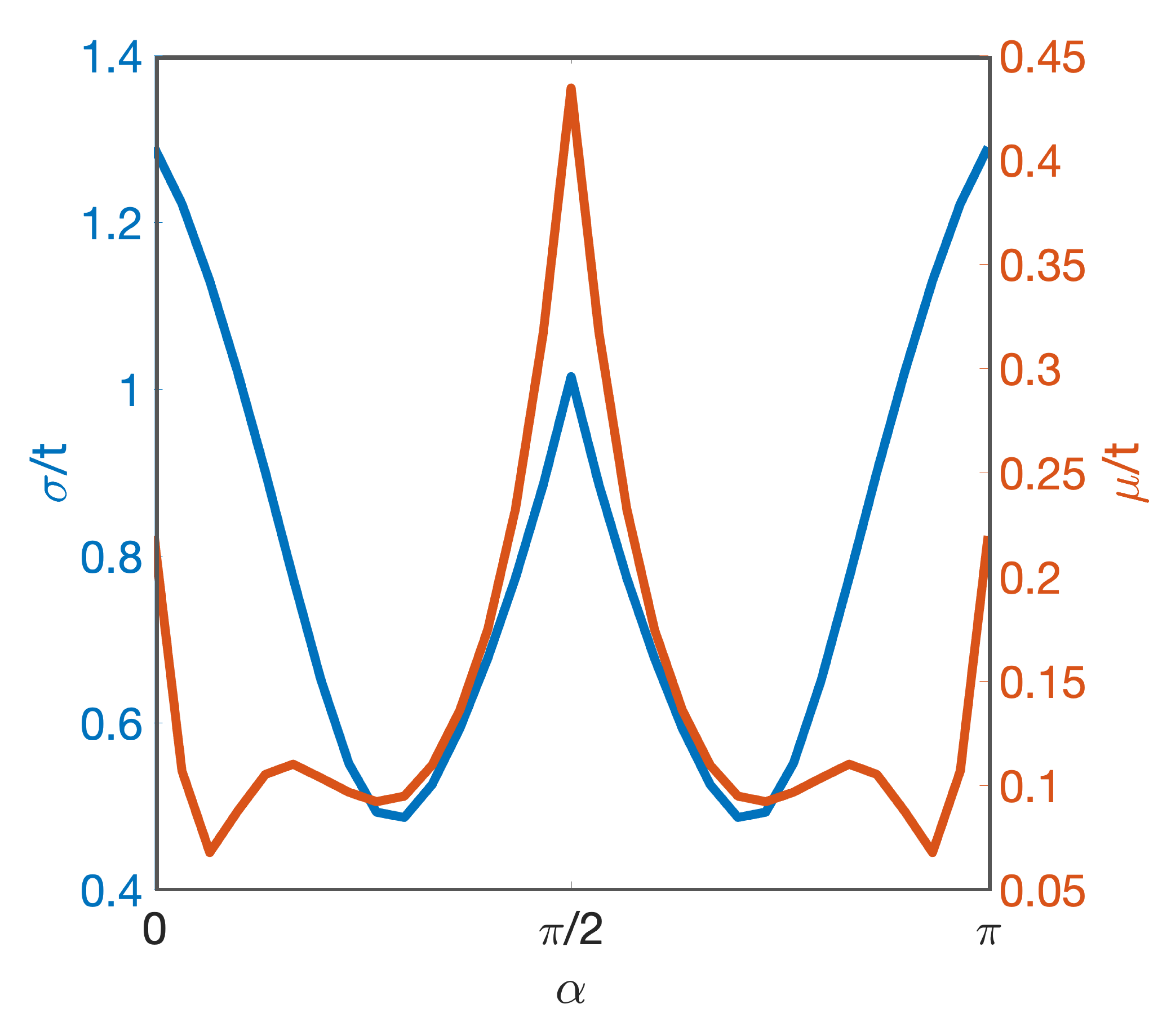}
\caption{In red, plot of $\mu/t$ and in blue plot of $\sigma/t$ as a function of $\alpha$.}
\label{fig:misigma}
\end{figure}

The evolution operator acting on the Hilbert space spanned by the vertices is
\begin{equation}
U=\sum_{k,\ell}\sum_{j,j'=0}^1\bracket{j'}{u}{j}\,\ket{\psi_{k\ell}^{j'}}\bra{\psi_{k\ell}^j}.
\end{equation}
The evolution operator analyzed in Ref.~\cite{PF17} is obtained if we take $\alpha=0$ and $\theta=\pi/4$. The evolution operator of this example is related to the evolution operator of the quantum walk analyzed in Ref.~\cite{FGGB11}.

\subsection{Transport}

In this section, we use again the first moment as the transport measure. The transport can be enhanced by choosing an appropriate localized initial condition and by tunning parameter $\alpha$. Fig.~\ref{fig:misigma} shows the mean $\mu=\sqrt{{\langle x \rangle}^2+{\langle y \rangle}^2}$ and the mean square displacement $\sigma=\sqrt{\sigma_x^2 +\sigma_y^2}$ as a function of $\alpha$ for the initial condition
\begin{equation}\label{ic-2d}
 \ket{\psi_0}=\frac{1}{2}\big(\ket{0,0}+\ket{1,0}+\ket{0,1}+\ket{1,1}\big).
\end{equation}
The maximum of the mean $\mu$ is obtained when $\alpha=\pi/2$. For this choice of $\alpha$, the mean is larger than the corresponding one for $\alpha=0$ and the mean square displacement (for $\alpha=\pi/2$) is smaller than the corresponding one for $\alpha=0$.

\begin{figure}[h!] 
\centering
\includegraphics[trim={900 60 900 60},scale=0.09]{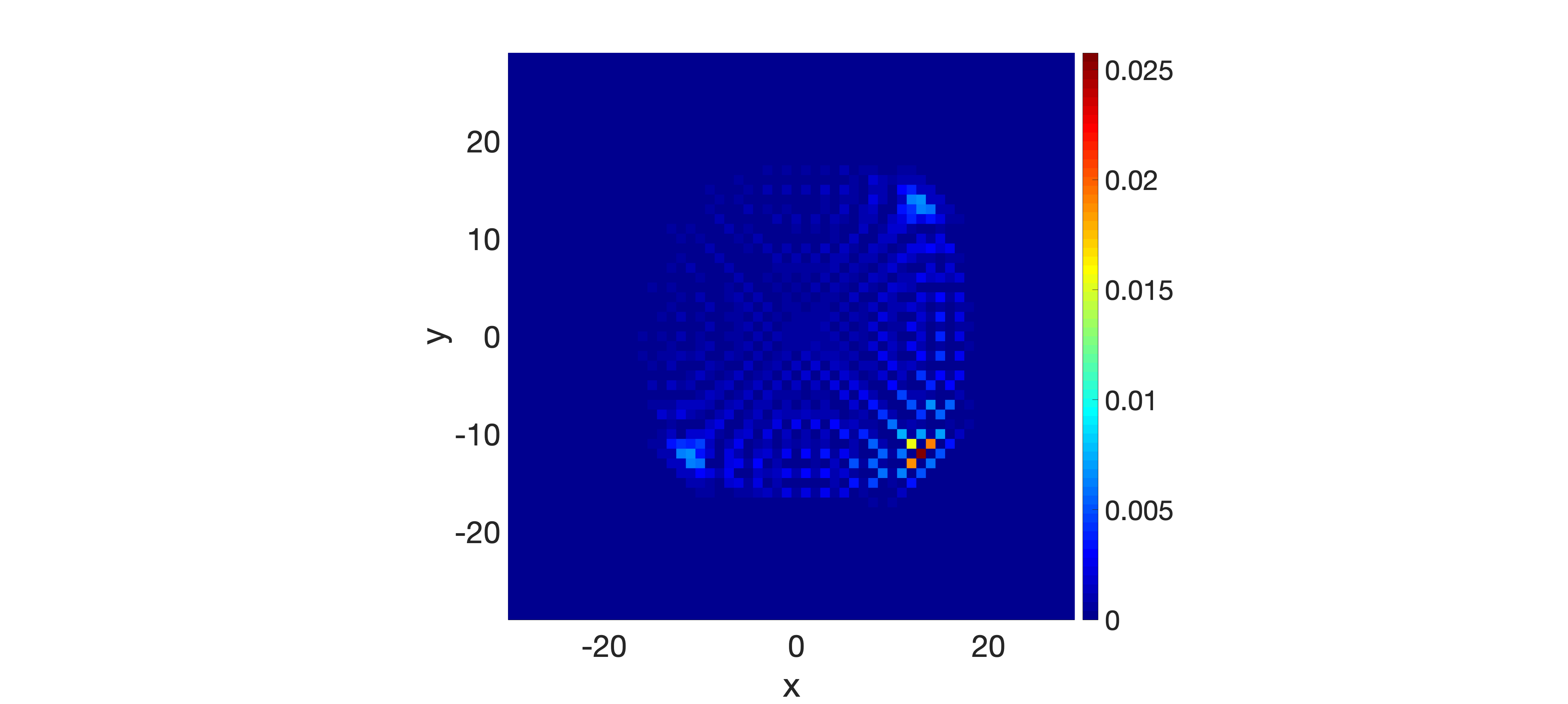}
\caption{Probability distribution after 13 steps of a oriented quantum walk with the same initial conditions of Fig.~\ref{fig:misigma}.}
\label{fig:probdist}
\end{figure}

Fig.~\ref{fig:probdist} show the probability distribution after 13 steps with $\alpha=\pi/2$ and the initial condition~(\ref{ic-2d}). Note that the probability distribution is concentrated around the lower corner and is as far as possible to the initial position.

\section{Conclusions}\label{sec:conc}

We have defined a discrete-time quantum walk on oriented graphs, whose evolution operator is the product of local operators obtained using partitions of the vertex set into cliques. In the discrete-time case, the evolution operator is the product of at least two local operators, because a single local operator traps the walker in cliques producing a trivial dynamics. In order to obtain interesting dynamics, we need to employ at least a second local operator. The minimum number of local \mbox{operators} depends on the graph structure. For instance, the two-dimensional lattice needs at least four local operators~\cite{ACFKMPP20}.

Using this newly defined quantum walk model on oriented graphs, we have analyzed the role played by parameter $\alpha$ on the quantum transport on the oriented line and oriented two-dimensional lattice. We have shown that the quantum transport can be enhanced or decreased by tuning parameter $\alpha$.

As a future work, we intend to analyze other transport measures in order to compare with the one we have used here. We also would like to check whether parameter $\alpha$ can be used to enhance quantum-walk-based search algorithms~\cite{Won15} and its connection with time-reversal symmetry breaking~\cite{ZFKWLB13}.

\section*{Acknowledgments}
We thank Fernando de Melo for insightful discussions and Takuya Machida for calling our attention to the equivalence between our second example and the alternate quantum walk. RP acknowledges financial support from CNPq grant n.~303406/2015-1 and Faperj grant CNE n.~E-26/202.872/2018.


\end{document}